\documentstyle{article}

\begin{document}

\newcommand{\fr}{\frac}
\newcommand{\non}{\nonumber \\ }
\newcommand{\oo}{\over }
\newcommand{\s}{\sqrt}
\newcommand{\be}{\begin{equation}}
\newcommand{\ee}{\end{equation}}
\newcommand{\bea}{\begin{eqnarray}}
\newcommand{\eea}{\end{eqnarray}}

\newcommand{\drm}{\delta\rho_m}
\newcommand{\drr}{\delta\rho_r}
\newcommand{\rmo}{\rho_{m0}}
\newcommand{\rro}{\rho_{r0}}

\newcommand{\al}{\alpha}
\newcommand{\e}{\eta}
\newcommand{\de}{\delta}
\newcommand{\r}{\rho}
\newcommand{\la}{\lambda}
\newcommand{\m}{\mu}
\newcommand{\n}{\nu}
\newcommand{\p}{\prime}
\newcommand{\G}{\Gamma}
\newcommand{\g}{\gamma}
\newcommand{\D}{\Delta}

\newcommand{\q}{\quad}
\newcommand{\qq}{\qquad}
\newcommand\ds{\displaystyle}

\newcommand{\ms}{\medskip}
\newcommand{\bs}{\bigskip}

\begin{titlepage}

\begin{center}

{\LARGE \bf Perturbations of a Universe Filled }
\ms
{\LARGE \bf with Dust and Radiation}

\ms
\bs
{\bf Z. Perj\'es}
\\
{\em KFKI Research Institute for Particle and Nuclear Physics, \\
H-1525 Budapest 114, P.O.Box 49, Hungary}
\\
\bs
{\bf A. Kom\'arik}
\\
{\em E\"otv\"os University, 1080 Budapest, R\'ak\'oczi £t 5, Hungary}

\bs
\today
\bs
\bs
\bs
\bs

\begin{abstract}

A first-order perturbation approach to $k=0$ Friedmann
cosmologies filled with dust and radiation is developed. Adopting
the coordinate gauge comoving with the perturbed matter, and
neglecting
the vorticity of the radiation, a pair of coupled equations is
obtained for the trace $h$ of the metric perturbations and for the
velocity potential $v$. A power series solution with upwards
cutoff exists such that the leading terms for large values of the
dimensionless time $\xi$ agree with the relatively growing terms
of the dust solution of Sachs and Wolfe.

\end{abstract}
\end{center}
\end{titlepage}

\section{Introduction}
  The classic prediction for the temperature fluctuations of the
cosmic background radiation by Sachs and Wolfe\cite{SW}
overestimates the experimental value\cite{Ma} by at least two
orders of magnitude. Worse than that, this prediction has been
obtained by neglecting the temperature fluctuations on the surface
of the last scattering, and including only the
gravitational perturbations along the subsequent path of the photon.
Obviously, if the initial fluctuations are random, their inclusion
can only increase the effect. While these computations can be
criticised on the basis that the results are not invariant with
respect to the choice of the initial hypersurface where the
photons originate\cite{Ellis}, this is not likely to be the way of
improving the results. Meanwhile, the discovery of large-scale
structures (such as voids or walls) certainly did not bring us
closer to the solution of this cosmic puzzle. It
is hard to take seriously
the suggestion\cite{Ma} that the observed
fluctuations are primordial (as opposed to propagation effects),
unless we are able to reduce
the magnitude of the Sachs-Wolfe estimate accordingly. The way the
surface of photon emission is defined is just one detail of the
gauge choice.  As we seek
diminuation of the effect, we may as well stick to the comoving
gauge. For gauge invariant treatments, {\it cf.} Refs.
\cite{Mag},\cite{Russ},\cite{KS2},\cite{Dun},\cite{Mukhanov}.

  Sachs and Wolfe have obtained, in a closed form, the first-order
perturbations of a Friedmann universe with a flat 3-space, and filled
either with dust or radiation.
They assumed that the domain of the Universe in which the
photon travels is matter-dominated, and
computed the temperature fluctuations in the dust-filled
universe. Their assumption is
justifiable because the decoupling occurs near the time of equal
matter-and-radiation density. (Some authors estimate that the two
phenomena occur simultaneously\cite{Wa} while others \cite{LS}
take that the equal-density epoch precedes the surface of last
scattering.)
According to the unperturbed models, the radiation density
$\r_r$
dies out faster than the matter density,
$\r_m\ .$
However, due to the instabilities, the dust dominance may not
hold everywhere in the perturbed models.

 Here we consider the refinement of the Sachs-Wolfe scheme by
computing the
perturbations of the $k=0$ Friedmann universe in the presence of
{\em both} dust and radiation. Instabilities are known to exist in
two-fluid cosmologies\cite{Mukhanov}. Assuming, for instance, that
the observed large structures are lately emerging manifestations
of the instabilities, one might be able to reduce the magnitude of
the metric fluctuations affecting the photon orbits.
The energy-momentum tensor is a sum of those of the two media,
\be
T^a_{\;b}  =T^{\,a}_{m\,b}+ T^{\,a}_{ r\,b}
\ee
The contribution of the dust has the form
\be
T^{\,a}_{m\,b}=- \r_m u^a u_b \ .
\ee
For the radiation,
\be
T^{\,a}_{r\,b}=-{4\oo 3} \r_r u^a u_b + {1\oo 3}\r_r \de^a_{\;b}
\ee
The four-velocities are normalized $u^au_a=1$.
We adopt the conformal form of the metric
\be
g_{ab}=a^2(\e) \left(\e_{ab}+h_{ab} \right) \label{g} ,
\ee
where the unperturbed ($h_{ab}=0$) metric satisfies\cite{MTW}
\be
 3 {1\oo a^2} {\left( {da \oo d\e }\right) }^2 -
 { {\r_m}_o {a_o}^3 \oo a} - { {\r_r}_o {a_o}^4 \oo a^2}=0\ .
\ee
such that the density $\r_i$ where $i$ stands either for $m$ (matter)
or $r$ (radiation), equals ${\r_m}_o$ or ${\r_m}_r$ at some
prescribed conformal time $\e=\e_0$.
The solution has the form

\be
a=\fr{1}{4}\la\e^2-\m
\ee
where the constants are defined

\be
  \la=\fr{1}{3}\rmo a_0^3    \qquad\qquad
  \m=\fr{\rro}{\rmo}a_0         \ .
\ee

\section{The perturbed model}
    In the perturbed universe, $h_{ab}\neq0$,
the densities of the components can be written to first order

\be
\r_i^{(1)} =\r_i+\de\!\r_i         \ .
\ee
Here $\r_r$ and $\r_m$ are the unperturbed densities

\be
\r_m(\e)= \rmo \fr{a_0^3}{a^3}
\qquad\qquad
\r_r(\e)=\rro \fr{a_0^4}{a^4}
\ee
and $\de\r_i$ are the first-order density perturbations. The
indices of the perturbed quantities are lowered and raised by the
Minkowski metric $\e_{ab}=\e^{ab}$.

    It can be proven\cite{lett} that a homogeneous universe cannot develop
perturbations with comoving dust and radiation. In fact,
the dipole effect of the cosmic background
radiation provides an experimental value for the local relative
velocity of the order of 100 km/sec. Thus we have good reason to
assume that $\de u^i_m\neq\de u^i_r$. We then choose
coordinates comoving with the matter:
\bea
u_m^a &=& u_{0}^a  \\
u_r^a &=& u_{0}^a+\de \!u_i^a
\eea
where the coincident unperturbed velocities are
\be  u_{0}^a={1\oo a}\de^a_0\ .\ee
The normalization conditions imply that $h_{00}=0$ and
$\de \!u_{r}^{0}=\;\de {u_r}_0 \;=0$.

  After the decoupling, we may assume that the conservation laws
apply separately both to the matter and the radiation components:
$T_{mb;a}^a=0$ and $T_{rb;a}^a=0$. Thus coupling occurs only via
the universal gravitational interaction. We get two energy
conservation equations from the timelike components and two
momentum conservations laws from the spacelike components.

\subsection{Energy conservation}

 The first-order perturbations of the {\em dust} satisfy the
equation
\be
{\left( { \de \! \r_m \oo \r_m} + {1\oo 2} h \right)}^{\p} = 0
\ee
where $h=h^\al_\al$ is the trace of the spacelike perturbation of
the metric and
a prime denotes partial derivative with respect to the
conformal time $\e$.
The solution is

\be
\de \! \r_m = \r_m \;\left( E ( x^{\beta} )- {1\oo 2} h \right)
\label{2d}
\ee
where the integration function $\;  E ( x^\beta )\;$ depends only
on the space coordinates $x^\al (\al=$1,2 or 3).

For the {\em radiation}, the energy conservation law has the form

\be
{\left(\; \r_r^{3\oo 4}\, \s g \, u^a_r \;\right)}_{,a} = 0
\ee
or
\be
{\left({3\oo 4}{\de \! \r_r \oo \r_r} + {1\oo 2} h \right)}^{\p}
+a \left( \de \!u^{\al}_r \right)_{,\al} =0  \label{er}
\ee

\subsection{Momentum conservation}

The conservation law for the {\em dust} has the form
\be
\left( a h_{0\al} \right)^{\p}=0
\ee
with the solution
\be
ah_{\al 0}=F_{\al}(x^\beta) \ .
\ee

\medskip
The remaining coordinate freedom makes it possible to arrange\cite{White}
\be
h^{0\al}_{\;\;\;,\al}=0 \ .
\ee

For {\em radiation}, the first-order momentum conservation law
reads
\be
\left( {\r_r}^{1/4} \de {u_r}_\al \right)^{\p}
={1\oo 4} a \r_r^{-3/4} \de {\r_r}_{,\al} \
.\label{mr} \ee

  The perturbed Einstein tensor will be written
\be
G_{\;b}^a={}_oG^a_{\;b} + \de G^a_{\;b}   \ .
\ee
Here ${}_oG^a_{\;b}$ is the unperturbed and $\de G^a_{\;b}$ the
first-order part. The Einstein equations for the first-order
quantities are
\bea
 \de G^0_{\;\;0}\; &=& -\left(\de \! \r_r + \de \! \r_m \right)    \\
 \de G^{\al}_{\;\;0}\; &=& - {4\oo 3} a \r_r \de \!u^{\al}_r    \\
 \de G^{\al}_{\;\;\beta}\; &=& {1\oo 3}\de^{\al}_{\;\,\beta} \de \! \r_r
\eea
Substitution of the Sachs-Wolfe\cite{SW} expressions for $\de
G^a_{\;b}$ and separating the trace-free part of the metric
perturbation
\be
S_{\al\beta}=h_{\al\beta}-{1\oo3}\e_{\al\beta}h
\ee
 yields
\be
 S^{\m\n}_{\quad ,\m\n}+ {2\oo 3} \D h
-2{a^{\p} \oo a} h^{\p}
\,=\; -\,2 a^2 \left(\de \! \r_r + \de \! \r_m \right)  \label{el}
\ee
\be
 S^{\al \m \quad \p}_{\quad ,\m}- {2\oo 3} {h^{,\al }}^{\p}+
\D h^{0 \al}-4 \left( \,2\, {{a^{\p}}^2 \oo a^2} \,
- \,{a^{\p \p} \oo a}\,\right)   \; h^{0 \al} \;=
- {4\oo 3} a \r_r \de \!u^{\al}_r \;,                 \label{ma}
\ee
\be
 2 h^{\p\p}+4 { a^{\p} \oo a }h^{\p} -
S^{\m\n}_{\quad ,\m\n}-{2\oo 3} \D h\;=\;-2a^2 \de \! \r_r   \ .\label{ha} \ee
\bea
{S^{\al}_{\;\beta}}^{\p\p}+2{ a^{\p} \oo a}{S^{\al}_{\;\beta}}^{\p}
 - \D S^{\al}_{\;\beta} &=&
S^{\al\m}_{\quad,\beta\m} +
S^{\quad\al\m}_{\beta\m,} - {2\oo 3} \de^{\al}_{\;\beta} S^{\m\n}_{\quad,\m\n}
 \non
&&  +h^{\al 0\quad \p}_{\quad ,\beta}
+ h_{\beta 0,}^{\quad\,\al\, \p}
+2{a^{\p} \oo a} \left( \>h^{\al 0\quad }_{\quad ,\beta}
+h_{\beta0,}^{\quad\,\al}  \>\right)     \non
&&  - {1\oo 3} h,_{\;\beta}^{\al}
- {1\oo 9}\de^{\al}_{\;\beta}\D h                    \label{ne}
\eea

 Taking the sum of
(\ref{el}) and (\ref{ha}) we get the simple relation

\be
h''+\fr{a'}{a}h'=-a^2 \left( 2 \drr+\drm \right) \ .\label{7}
\ee

\section{The velocity potential}
 For economy of writing, we introduce the scaled radiation
velocity perturbation
\be
v_{\al} (x_{\al},\e)=\r_r^{1/4} \de \!u_{r \al} \ .
\ee
(Note that the component $v_0=0$.)
{}From the $\e$ derivative
of (\ref{er}) and eliminating the term $a^2 \left( \de
\!u^{\beta}_r\right)$ by use of the radiation momentum
conservation (\ref{mr}) we get the relation
\be
\left[\r_r^{1/4}a \left(\fr{3}{4}\fr{\drr}{\r_r}+\fr{1}{2}h\right)' \right]'
+\fr{1}{4} a \r_r^{-3/4}(\drr)^{,\al}_{\;\;,\al}=0 \ .\label{l}
 \ee
Taking the time derivative and the gradient, and using again (\ref{mr})
to get rid of the divergence term, the left-hand side is still a
total time derivative. Integration and comparison with (\ref{l})
leads to the uncoupled equation
\be
v_{\beta\;\;,\al}^{\;\;,\al}-v_{\;\;,\al\beta}^\al=Q_\beta(x)
\label{l2}
\ee
where the integration function $Q_\beta(x)$ depends only on space
coordinates.
We decompose the three-vector $v_\al$
 \be
v_{\al}=\omega_{\al}+v,_{\al}
\ee
where the {\em vorticity} $\omega_{\al}$ is divergenceless
\be
\omega_{\al},^{\al}=0
\ee
and  $v$ is the {\em velocity potential}.
The terms containing the velocity potential $v$ cancel in Eq.
(\ref{l2}). Thus we get the simple relation for the vorticity
\be
\Delta\omega_{\beta}=-Q_\beta(x)\ .
\ee
Following a suggestion of \cite{Liddle},
we shall henceforth take the trivial solution $\omega_{\al}=0$
and $Q_\beta=0$ such that we have
\be
v_{\al}=v,_{\al} \ .
\ee

Using the velocity potential in Eq. (\ref{mr}), we get
\begin{equation}
 v_{,\alpha}' = {1\oo  4}a \rho_r^{-3/4} \delta\rho_{r,\alpha}\ .
\end{equation}
This can be integrated.
Since only the gradient of the velocity potential has
physical meaning, the integration function $U(\eta)$
may be chosen to vanish. Thus
\begin{equation}
v' = {1\oo  4}a \rho_r^{-3/4} \delta\rho_{r}\ .\label{123}
\end{equation}
Hence, Eq. (\ref{er}) can be rewritten
\be
3v''-\Delta v+ {1\oo 2}\rho_{r0}^{1/4}a_0 h' =0 \ . \label{v0}
\ee

Introducing the velocity potential (\ref{123}) in (\ref{7}), we
get
\begin{equation}
h''+{a'\oo  a}h'+\rho_{m0}{a_{0}^{3}\oo  a}\left(E(x)-{1\oo
2}h \right)+8a\rho_{r}^{3/4}v' =0 \label{f0}
\end{equation}
We may get rid of the inhomogeneous term $E(x)$ by introducing
the function
\begin{equation}
f=h-2E(x)   \ .
\end{equation}
The explicit form of the scale factor $a$ can be made simpler by
use of the dimensionless time variable
\begin{equation}
\xi={1\oo 2}\sqrt{\lambda\oo  \mu}\eta \ .
\end{equation}
Then we have
\begin{equation}
a=\mu(\xi^2-1) \hskip 3cm \left(\;\; \mu={\rho_{r0}\oo
\rho_{m0}}a_0 \;\;\right) \end{equation}
such that the Big Bang occurs at $\xi=1$.
Eqs. (\ref{v0}) and (\ref{f0}) take the form
\bea
\dot{v}-L\Delta v+{16\oo  K} \dot{f} &=&0 \label{v}\\
(\xi^2-1)\ddot{f}+2\xi \dot{f}-6f+K{1\oo  \xi^2-1}\dot{v}
&=&0,\label{f}
\eea
where an overdot denotes derivation with respect to the
dimensionless time $\xi$ and
the constants $K$ and $L$ are defined by
\bea
K&=&16 \s 3 \rho_{m0}\rho_{r0}^{-3/4} \\
L&=&4{\rho_{r0}\oo  \rho_{m0}^2}{1\oo  a_0^2} \ .
\eea

  These coupled linear equations for $f$ and $v$ have an elaborate
structure, even in the asymptotic regime.
In the special case of the velocity potential being stationary,
$\dot v=0$, Eq. (\ref{f}) becomes the $n=2$ Legendre equation
for $f$. In the generic case, however, the expansion in Legendre
series is cumbersome.

\section{Recursion relations}
The power series expansion
in the time variable $\xi$,
\bea
f(x,\xi)&=&\sum_{n=-\infty}^{\infty} a_n(x)\xi^n \\
v(x,\xi)&=&\sum_{n=-\infty}^{\infty} b_n(x)\xi^n
\eea
with Eqs. (\ref{v}) and (\ref{f}),
yields the recursion relations for the coefficients:
\bea
(n+1)(n+2)b_{n+2}-L\Delta b_n+
{16\oo K}(n+1)a_{n+1}&=&0 \label{136} \\
(n+1)(n-4)a_{n-2}-(2n^2-6)a_n  & &
\nonumber  \\
+(n+1)(n+2)a_{n+2}+K(n+1)b_{n+1}&=&0 \label{137}
\eea

  Eliminating $a_n$ we get
\bea
(n+1)(n-1)(n-4)b_{n-1}-(n+1)(2n^2+10)b_{n+1}  &&
\nonumber \nopagebreak  \\
+(n+1)(n+2)(n+3)b_{n+3}-{L\oo  n-2}(n+1)(n-4)\Delta b_{n-3} &&
\nonumber \\
+L{(2n^2-6)\oo  n}\Delta b_{n-1}- L(n+1)\Delta b_{n+1} &=& 0 \ .
\label{138} \eea

The convergence of the series expansion deserves further
investigation.
The terms with negative exponents will decay with time, and
it is expected that their
contribution to the temperature fluctuations becomes negligible.
Hence, to evade fast growing terms, it appears worth while to seek
solutions whith a maximal value of the exponent $n$.

  It is clear from the structure of Eqs. (\ref{136}) and (\ref{137})
that a cutoff of the power series can occur only simultaneously
in $a_n$ and $b_n$. Further
necessary conditions for cutoff may be
obtained from Eq. (\ref{138}).
The highest possible nonvanishing
coefficient is $b_1$. From (\ref{136}) it then follows that
$a_2$ is the highest coefficient of $f$. It is impossible to have
a cutoff simultaneously both upwards and downwards.
Our choice of the velocity potential implies that
any coefficient $b_n$ satisfying $\D b_n=0$ must vanish.
Thus the leading terms of the solution of Eqs. (\ref{136}) and
(\ref{136})  are

\bea
h(x,\xi)&=&a_2(x)\xi^2+\left(2E-{1\oo 3}a_2(x)\right)+a_{-2}(x){1\oo \xi^2}
  \nonumber \\
  \nopagebreak
        & &+a_{-3}(x){1\oo \xi^3}+{1\oo 3}a_{-2}(x){1\oo \xi^4}+... \\
v(x,\xi)&=&b_1(x)\xi+b_{-3}(x){1\oo \xi^3}+b_{-4}(x){1\oo \xi^4}+...
\eea
where the coefficients $a_n$ and $b_n$ satisfy the relations

\bs
$
\begin{array}{lll}
& a_0(x)=-{1\oo 3}a_2        & \Delta b_1(x)={32\oo KL}a_2(x)  \\
& a_{-2}(x)={K\oo 4}b_1(x)   & \Delta b_{-3}(x)={8\oo L}b_1  \\
& a_{-4}(x)={a_{-2}(x)\oo3}  & \Delta b_{-4}(x)={48\oo KL}a_{-3}(x) \ .

\end{array}
$
\bs

The leading terms in the density contrasts are
\begin{equation}
{\drr \oo \rho_r}=4\rro^{-1/4}a_0^{-1}\left( b_1(x)-3b_{-3}(x){1\oo \xi^4}-
4b_{-4}(x){1\oo \xi^5}+...\right)
\end{equation}
\begin{equation}
{\drm \oo \rho_m}=-{1\oo 2}\left(a_2(x)\xi^2+a_0(x)+a_{-2}(x){1\oo \xi^2}+
a_{-3}(x){1\oo \xi^3}+...\right) \ .
\end{equation}

  Thus we establish the comforting result that for large values
of $\xi$, the
leading power in time of the metric function $h$ coincides with
the power of the relatively growing mode of the pure dust
solution\cite{SW}. For a more detailed account of our approach,
{\it cf.} \cite{Kom}.

{\bf Acknowledgement}

The authors thank Professor J. Ehlers for having suggested this
research problem.

\end{document}